# External Cavity 637-nm Laser with Increased RSOA-to-PIC Alignment Tolerance and a Filtered Sagnac-Loop Reflector with Single Output Waveguide

Georgios Sinatkas,[1][0000-0003-4811-9601] Arijit Misra,[1, 2][0000-0002-0284-9708] Florian Merget,[1][0000-0001-8208-1681] and Jeremy Witzens[1][0000-0002-2896-7243]

[1] Institute of Integrated Photonics, RWTH Aachen University, Aachen 52074, Germany
[2] Current affiliation: Cisco Optical GmbH, Nordostpark, Nuremberg 90411, Germany
gsinatkas@iph.rwth-aachen.de

**Abstract.** The design of a 637-nm wavelength, photonic-integrated-circuit-based external cavity laser (PIC-based ECL) aimed at quantum technology applications is presented together with first experimental results. The PIC is designed to provide relaxed alignment tolerance for coupling to a reflective semiconductor optical amplifier (RSOA) gain chip. This is achieved by using a multi-mode edge coupler (MMEC) in place of the usually employed single-mode coupling schemes. A 1-dB-penalty misalignment tolerance of up to ±2.4 μm can be achieved in the plane of the chip, creating a path towards reliable flip-chip integration at short wavelengths. The power coupled to the PIC is fed to a Sagnac-loop reflector, filtered by a pair of ring resonators operated in Vernier configuration for providing the required frequency selective optical feedback. The ring resonators are designed to have different loaded $Q$-factors and they are asymmetrically coupled to bus and drop waveguides with suitably engineered directional couplers to provide single output waveguide emission. Moreover, requirements for high output power and narrow linewidths are balanced. Finally, preliminary measurements strongly suggest lasing in the fabricated devices, with further performance optimization being currently carried out.

## 1    Introduction

Photonic-integrated-circuit-based external cavity lasers (PIC-based ECLs) are attracting increased attention due to their reduced footprint, narrow linewidth, and tunability, which make them ideal sources for on-chip applications. Several PIC-based ECLs have been demonstrated in the infrared regime [1, 2, 3] for use in coherent communications, sensing, and metrology. However, there is a growing interest for similar demonstrations in the visible regime [4, 5] to serve the emerging fields of bio-photonics and quantum technology. In this work, a PIC-based ECL at 637 nm is presented, that is intended for exciting nitrogen vacancy centers in diamond, one of the most promising platforms for quantum information processing. In the following sections, the design of the PIC is presented together with preliminary ECL measurements.

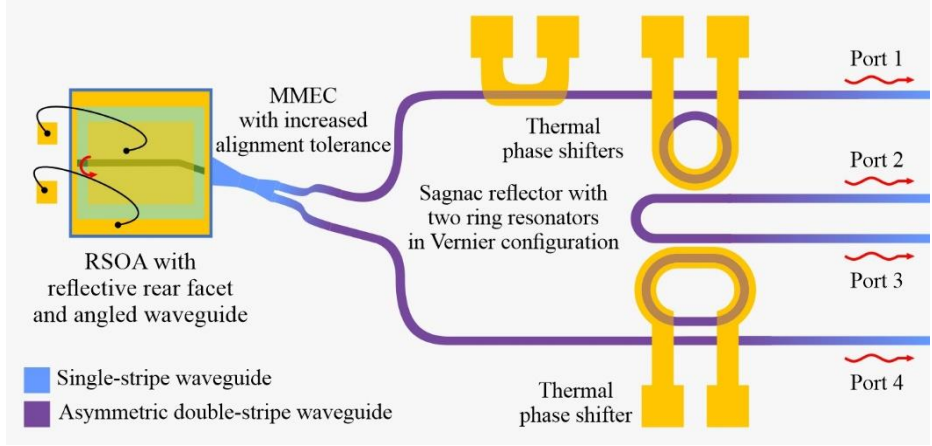

**Fig. 1.** Schematic representation of the ECL architecture.

## 2   Design

With reference to Fig. 1, the ECL is configured as a Fabry–Pérot cavity between the broadband, highly reflective back facet of an RSOA and a frequency selective reflector on the PIC. A highly misalignment-tolerant multi-mode edge coupler (MMEC) is used for the RSOA-to-PIC coupling, designed to equally split the coupled power between two single-mode on-chip waveguides, irrespectively of the in-plane misalignment between the two chips [6]. These are looped back onto each other to form a Sagnac loop, which provides a filtered reflection by interposing a pair of thermally tuned ring resonators with slightly different free-spectral ranges (FSRs), configured in an add-drop structure. By spectrally aligning their resonances, a broader effective FSR is achieved (Vernier effect), resulting in a single resonance in the gain bandwidth of the RSOA, facilitating single-mode lasing operation. The ring resonators are asymmetrically side coupled by suitably configured directional couplers, having, thus, different loaded $Q$-factors [7], which allow for single-output operation and for balancing between high-output-power and narrow-linewidth requirements.

### 2.1   Silicon Nitride Platform

Silicon-nitride (SiN) technology is selected due to its low absorption at visible wavelengths. Two waveguide types are employed: a single-stripe (SS) and an asymmetric double-stripe (ADS) geometry, embedded in a silicon dioxide ($SiO_2$) cladding [Fig. 2(a), (b)]. The thickness of the bottom SiN stripe is set equal to 27 nm to achieve optimal overlap between the RSOA-beam profile and the mode profile of the MMEC in the vertical direction (Section 2.2). The top SiN layer and the intermediate $SiO_2$ have thicknesses of 80 nm and 50 nm, respectively. The etching angle $a_g$ equals 82°. The components were designed into the process of LioniX International [8].

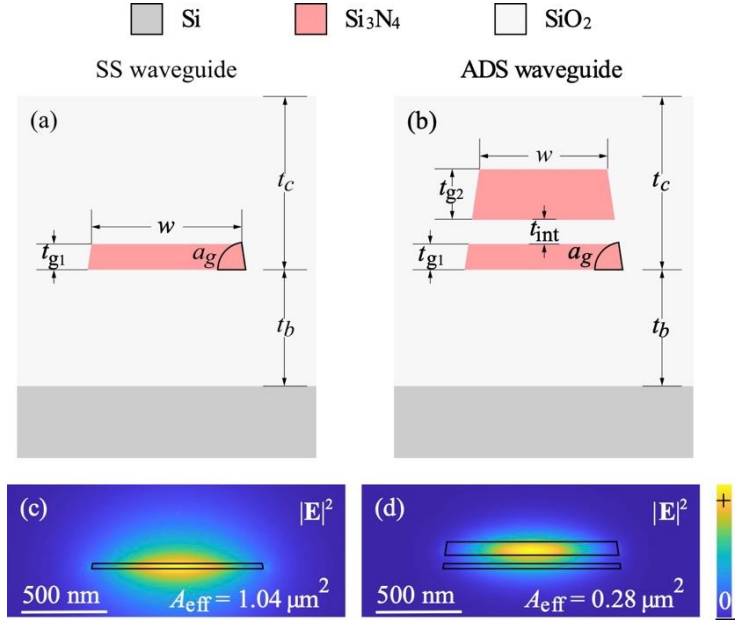

**Fig. 2.** (a) SS and (b) ADS waveguide with $t_{g1} = 27$ nm, $t_{int} = 50$ nm, $t_{g2} = 80$ nm, $a_g = 82°$, $t_c = 4$ μm, $t_b = 8$ μm. (c), (d) Respective $TE_0$ mode.

The weak confinement of the SS waveguide is suitable for matching the large beam profile of RSOAs and optical fibers [Fig. 2(a)], while the ADS waveguide offers tighter confinement [Fig. 2(b)] and it is favorable for bends and ring resonators. The transition between the two waveguide types is achieved using vertical tapers, interposed at the two (four) output ports of the MMEC (chip) [Fig. 1].

Waveguide loss stems mainly from scattering at the side SiN waveguide walls due to roughness after etching. To mitigate this effect, the interconnect waveguide width is selected wide enough to minimize the overlap of the mode with the side walls, while maintaining single-mode operation. Based on the dispersion diagrams in Fig. 3 and considering TE operation, the single-mode waveguide width for the SS (ADS) waveguide is selected equal to 1.5 μm (0.55 μm). In the MMEC, the narrowest width is set to 2.5 μm, which supports both $TE_0$ and $TE_1$ modes as required for its operation.

### 2.2 Components

**Alignment Tolerant Multi-Mode Edge Coupler.** Compared to its single-mode counterparts, the MMEC relaxes the requirements for alignment in the plane of the PIC. This is particularly useful when a complete PIC-RSOA integration is attempted, using, e.g., the flip-chip bonding technique with vertical self-alignment [9, 10]. The MMEC is designed into the SS waveguide, considering the beam characteristics of the RSOA (EXALOS AG). The latter has an elliptical profile with a mode field diameter (MFD) equal to 3.85 μm (0.82 μm) in the horizontal (vertical) direction [Fig. 4(a)], which is optimally matched by the $TE_0$ mode of an SS waveguide with a thickness of 27 nm and a width of 4.5 μm [Fig. 4(b)].

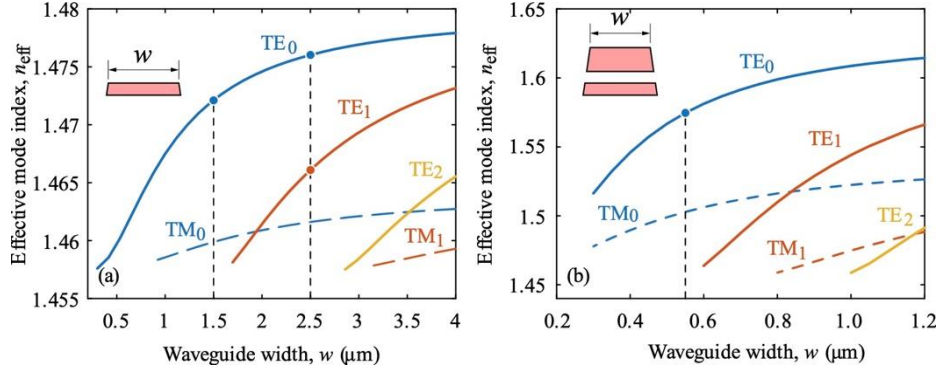

**Fig. 3.** Mode dispersion diagrams for the (a) SS and (b) ADS waveguide as a function of width.

In case of misalignment [Fig. 4c)], part of the emitted RSOA power couples to higher order modes due to the large MMEC width. Fig. 4(d) shows the increase in the horizontal misalignment tolerance from ±0.8 μm to ±1.5 μm, defined at 1-dB drop from the peak coupling-efficiency, obtained if only the power coupled to the next higher order mode ($TE_1$) is recovered. This is achieved by incorporating a 1x2 multi-mode interference (MMI) splitter as part of the MMEC (Fig. 5), with a length that ensures that the $TE_0$ and $TE_1$ modes reach the output waveguides in quadrature to equally split the power. The narrowest $L_2$-long section of the MMEC is chosen to filter out modes with orders higher than $TE_1$. The misalignment tolerance can be further increased to ±2.4 μm for a 7-μm wide MMEC [Fig. 4(d)], with a slight IL penalty occurring at $\Delta x = 0$, but equally high peak coupling efficiency at slight lateral displacements. The calculations in Fig. 4(d) are performed using overlap integrals between the $TE_0$, $TE_1$ waveguide modes and the RSOA beam profile. The two designs of the MMEC are listed in Table 1.

**Table 1.** MMEC designs with different 1-dB misalignment tolerance (units: microns).

| Design | $w_1$ | $w_2$ | $L_1$ | $L_2$ | $L_3$ | $L_4$ | $L_5$ | $s$ | Tolerance |
|---|---|---|---|---|---|---|---|---|---|
| #1 | 4.5 | 4.0 | 40 | 29.273 | 30 | 6 | 100 | 2 | ±1.5 |
| #2 | 7.0 | 4.0 | 50 | 16.012 | 30 | 6 | | | ±2.4 |

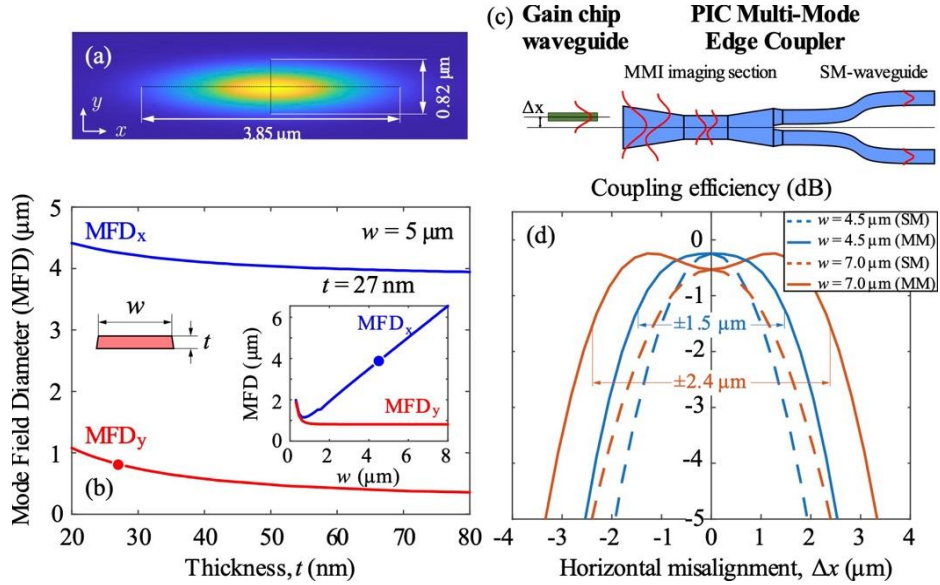

**Fig. 4.** (a) RSOA beam and near-field MFDs in the horizontal and vertical direction. (b) MFD for the TE$_0$ mode of the SS waveguide as a function of thickness and width (inset). (c) Schematic of the RSOA-to-PIC coupling for horizontal misalignment $\Delta x$. (d) Coupling efficiency between the RSOA beam and the TE$_0$, TE$_1$ modes of a 4.5-μm and 7-μm wide SS edge coupler as a function of $\Delta x$.

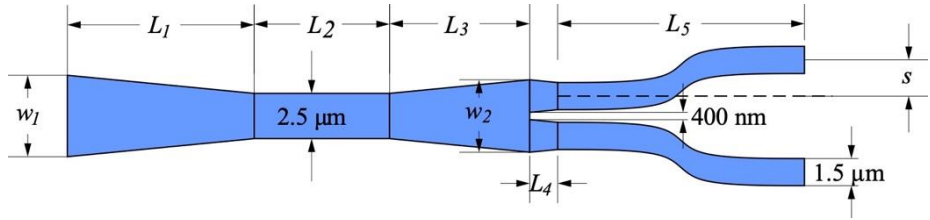

**Fig. 5.** Schematic of the MMI splitter with all pertinent dimensions labeled.

**Ring-Resonator-based Sagnac-Loop Reflector.** This component is responsible for providing the required frequency selective feedback. It employs the ADS waveguide due to its tighter confinement, allowing for smaller bend radii and, thus, higher FSR. Setting the limit of maximum bend loss to 0.01 dB/cm, which is two orders of magnitude lower than the expected propagation loss (1 dB/cm), a minimum radius of 25 μm is considered for the 550 nm wide waveguide [Fig. 6(a)], resulting in an FSR ≈ 1.42 nm and approximately four resonances in the 6-nm RSOA gain bandwidth.

To suppress the extra resonances, the Vernier effect is employed in two serially cascaded rings with slightly different radii and FSRs (1.41 nm and 1.31 nm). By tuning the ring resonances to the desired wavelength using thermal phase shifters [Fig. 6(b)], the effective FSR of the total system increases to $FSR_A \times FSR_B/|FSR_A - FSR_B|$ = 17 nm, which is wide enough to allow for a single resonance in the gain spectrum of the RSOA.

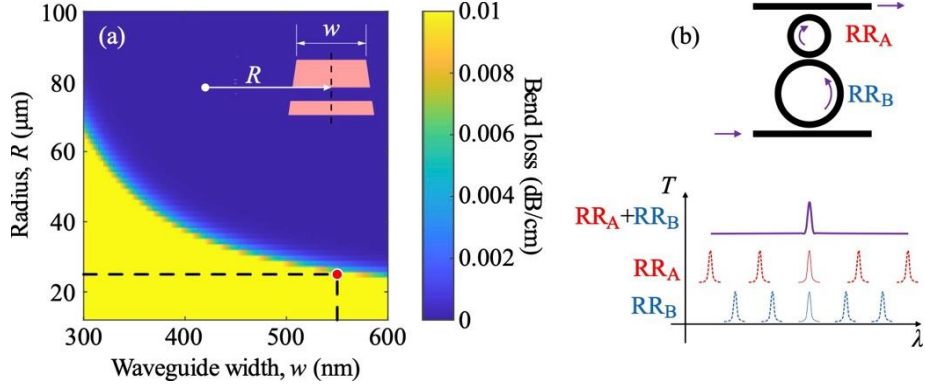

**Fig. 6.** (a) Bend losses for the ADS waveguide as a function of its width and bend radius. (b) Vernier effect in a pair of rings with slightly different FSRs in cascade configuration. The resulting add-drop filter features a wider FSR.

In the literature, identical directional couplers are usually employed. In this work, an asymmetric design is suggested, optimally engineered to route the majority of the output power to a single output port. Such a design is expected to minimize undesired back-reflections from the waveguide ends that could destabilize the laser operation. In addition, it facilitates the integration of the PIC with fiber optics and isolators.

For the design, the circuit model representation of the add-drop filter is employed [11]. With reference to Fig. 7(a), Port 1 is selected as the nominal output port. The power exiting Port 4 [Fig. 7(b)] can be theoretically zeroed out if the transmission amplitude coefficients $t_{B4}$, $t_{B3}$ for the respective directional couplers with gaps $g_{B4}$, $g_{B3}$ satisfy the critical-coupling condition [11],

$$t_{B4} = a_B \, t_{B3}, \qquad (1)$$

with $a_B$ the round-trip amplitude attenuation for $RR_B$. Considering lossless couplers, Eq. (1) can be equivalently expressed using the respective coupling coefficients as

$$\kappa_{B4} = \sqrt{1 - (1 - \kappa_{B3}^2)a_B}. \qquad (2)$$

Similarly, the power reaching Port 2 in Fig. 7(b) can be fully suppressed as long as

$$\kappa_{A2} = \sqrt{1 - (1 - \kappa_{A1}^2)a_A}. \qquad (3)$$

Consequently, the power coupled to Ports 1 and 3 as well as the total optical feedback are rendered functions of $\kappa_{A1}$ and $\kappa_{B3}$.

This still leaves the issue of minimizing the power being coupled to Port 3 [Fig. 7(a)]. This is addressed by designing $RR_B$ to have a much lower loaded $Q$-factor (~ 31K) than $RR_A$ (~ 300K). Providing $RR_B$ with high waveguide coupling coefficients such that the intrinsic losses play a lesser role, on the one hand, allows for both coupled waveguides to experience close-to-critical coupling, minimizing the power lost to Port 3. Providing $RR_A$ with a high $Q$-factor, on the other hand, allows for better filtering and lower linewidth operation.

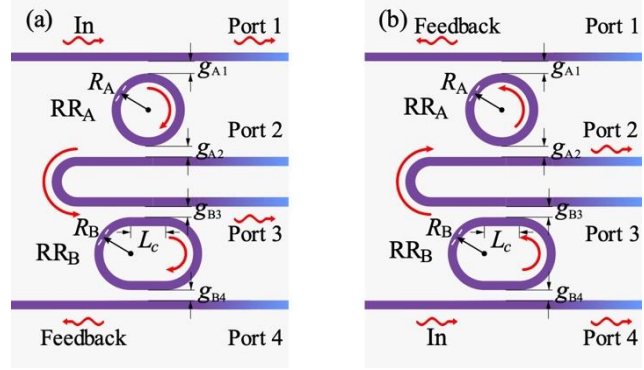

**Fig. 7.** Wave propagation through the ring-resonator-based Sagnac-loop reflector in the two possible directions.

As expected, the numerical optimization in Fig. 8(a) and (b) confirms that the power reaching Port 1 (Port 3) becomes maximum (minimum) for low $\kappa_{A1}$ (high $\kappa_{B3}$) values. The on-resonance transmission to Port $N = \{1, 3\}$ is reduced to the analytical form

$$T_N = \left( \frac{a_X^2 - 1}{t_{XN} a_X^2 - \frac{1}{t_{XN}}} \right)^2, \qquad (4)$$

with $X = \{A, B\}$ indicating the respective side-coupled resonator, after substituting Eqs. (2) and (3) in Eq. (5) in [11]. Fixing $\kappa_{B3}$ at 0.2, the coupling coefficient $\kappa_{A1}$ can be varied to achieve the required balance between output power [Fig. 8(a)] and optical feedback [Fig. 8(c)] or, equivalently, ECL linewidth [Fig. 8(d)]. The latter is estimated using the formalism described in [12, 13]. After optimizing all four coupling coefficients, suitable directional-coupler geometries in Fig. 7 can be obtained through full-wave simulations.

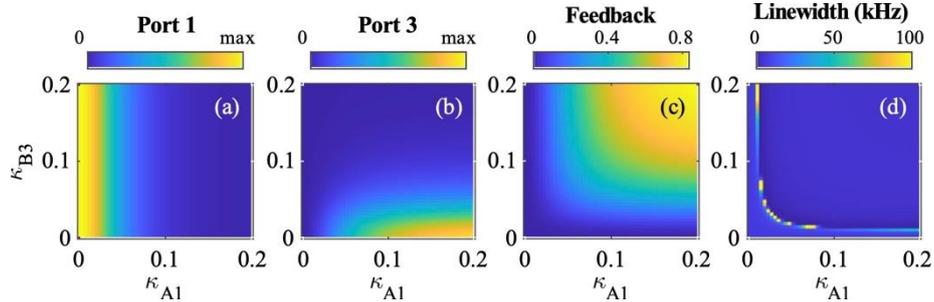

**Fig. 8.** Change in the power at (a) Port 1 and (b) Port 3, (c) optical feedback, and (d) theoretical ECL linewidth as a function of the coupling coefficients $\kappa_{A1}$ and $\kappa_{B3}$.

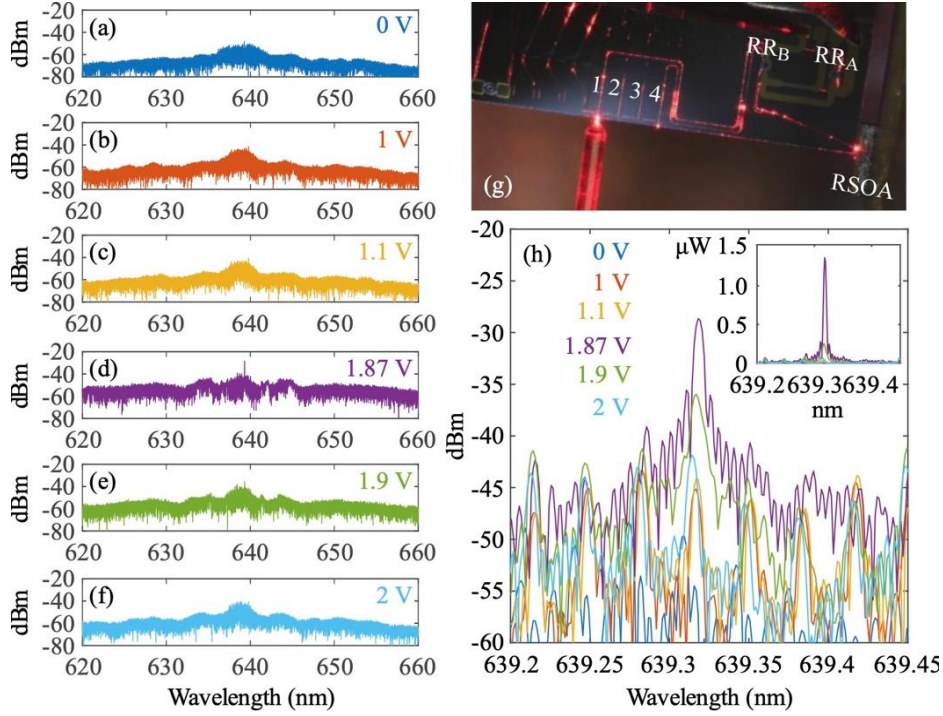

**Fig. 9.** (a)-(f) Spectrum evolution for six different biasing of RR$_B$, with RR$_A$ remaining unbiased. At 1.87 V, the two ring resonances overlap and a distinctive peak is evidenced in the emission. (g) Snapshot of the hybrid RSOA-to-PIC coupling. (h) Close-up view of the spectra in (a)-(f). Inset: in linear scale.

## 3    Experimental Results

In Fig. 9, preliminary measurements from the hybrid RSOA-to-PIC setup [Fig. 9(g)] are presented. The PIC under test has a 7-μm wide MMEC (Table 1) and is coupled to an RSOA mounted on an alignment stage. The ROSA is biased at 78 mA with an output power of ~7 mW. RR$_A$ is a perfect ring ($R_A$ = 24.97 μm), while RR$_B$ has a racetrack geometry, with $R_B$ = 23.78 μm and $L_c$ = 10 μm. The gap values equal $g_{A1}$ = 541 nm, $g_{A2}$ = 425 nm, $g_{B3}$ = 421 nm, $g_{B4}$ = 415 nm, while the calculated $\kappa$ values are 0.0326, 0.0687, 0.21, 0.2188, respectively. By simply biasing RR$_B$ while keeping RR$_A$ unbiased, the resonance of the former is redshifted until both resonances overlap and a distinctive light-peak manifests itself [Fig. 9(d)]. This peak cannot be attributed to filtered amplified spontaneous emission (ASE) since the latter does not have to pass through the

Sagnac loop to reach Port 1. An increase in bias shifts the $RR_B$ resonance further, beyond $RR_A$, and the effect ceases. From Fig. 9(h), the peak center is measured at 639.32 nm with a linewidth of 4.93 pm, limited by the 10 pm resolution bandwidth of the optical spectrum analyzer (Thorlabs OSA201C). More accurate linewidth measurements will be performed using, e.g., delayed self-heterodyne interferometric detection [14]. The presented preliminary results strongly suggest the manifestation of lasing, with many degrees of freedom being still available for optimizing the performance.

## 4   Conclusions

The design of a PIC-based ECDL at 637 nm was presented, including an MMEC with relaxed alignment tolerance and an asymmetric ring-resonator-based Sagnac-loop reflector, designed for achieving single-output operation and flexible optimization between output power and linewidth. Preliminary measurements strongly indicate lasing, with a linewidth value limited by the OSA resolution.

**Acknowledgments** Funded by the Federal Ministry of Education and Research (BMBF) - 13N15965.